\documentclass{osa-article}

\usepackage{xcolor}

\usepackage{siunitx}
\usepackage{physics}
\journal{osajournal}
\articletype{Research Article}
\begin{document}

\title{Ultra-Thin All-Epitaxial Plasmonic Detectors}

\author{Leland Nordin,\authormark{1} Priyanka Petluru,\authormark{1} Abhilasha Kamboj,\authormark{1} Aaron J. Muhowski,\authormark{1}  and Daniel Wasserman\authormark{1,*}}

\address{\authormark{1}Department of Electrical and Computer Engineering, University of Texas at Austin, Austin, TX 78758 USA\\}

\email{\authormark{*}dw@utexas.edu} %% email address is required



\begin{abstract*}
{We present an infrared photodetector leveraging an all-epitaxial device architecture consisting of a ‘designer’ plasmonic metal integrated with a quantum-engineered detector structure, all in a mature III-V semiconductor material system.  Incident light is coupled into surface plasmon-polariton modes at the detector/‘designer’ metal interface, and the strong confinement of these modes allows for a sub-diffractive ($\sim \lambda_0 / 33$) detector absorber layer thickness, effectively decoupling the detector's absorption efficiency and dark current. We demonstrate high-performance detectors operating at non-cryogenic temperatures (T = 195 K), without sacrificing external quantum efficiency, and superior to well established and commercially-available detectors.  This work provides a practical and scalable plasmonic optoelectronic device architecture with real world mid-infrared applications.}
\end{abstract*}

%%%%%%%%%%%%%%%%%%%%%%%%%%  body  %%%%%%%%%%%%%%%%%%%%%%%%%%
\section*{Introduction}

The field of nanophotonics has been driven by the desire to demonstrate optical structures capable of sub-diffractive confinement of light, and to subsequently leverage these structures to develop optoelectronic devices capable of emitting, detecting and/or manipulating light at these same sub-diffractive length scales. The field of plasmonics has long been central to such efforts, as the strong optical confinement provided by plasmonic structures offers a path towards sub-diffraction-limit optical waveguides, sources, and sensors\cite{Maier2003LocalWaveguides,Anker2008BiosensingNanosensors,Atwater2010PlasmonicsDevices,Schuller2010,Noginov2009DemonstrationNanolaser, Oulton2009PlasmonScale,Hill2009,Nakayama2008PlasmonicCells,Nikolajsen2004SurfaceWavelengths}. However, while plasmonic devices can offer impressive mode confinement, they often underperform diffraction-limited devices on most other metrics\cite{Khurgin2015HowMetamaterials}. The prospects for scalable plasmonic architectures for real world applications have thus largely been limited, and plasmonics has had its greatest impact in applications where the strong optical mode confinement benefits all-optical sensing of small material volumes \cite{Kneipp1997,Nie1997} or localized heat generation \cite{Halas2014,Drezek05}, neither of which are true optoelectronic realizations of plasmonic enhancement.  Transitioning plasmonics from optical structures\cite{Okamoto2004} to optoelectronic devices\cite{Kwon2008SurfacePlasmonDiode} has not, until now, resulted in performance competitive with existing diffraction-limited technologies.

For the longer wavelengths of the mid-infrared (mid-IR, $3-\SI{20}{\um}$) and particularly in the long-wave infrared (LWIR, $8-\SI{13}{\um}$), achieving nano-scale devices is substantially more challenging than at the shorter wavelengths where the bulk of plasmonics research efforts are undertaken\cite{Stanley2012PlasmonicsMid-infrared}.  This is partially a result of the order of magnitude increase in wavelength (and proportionally, the diffraction limit) as one moves from the near-infrared/visible (near-IR/vis) wavelength range to the LWIR but also because the traditional plasmonic metals of the near-IR/vis, with increasing wavelength, behave more and more like perfect electrical conductors, and are unable to support highly confined plasmonic modes\cite{Law2013TowardsPlasmonics}.  However, in the mid-IR, highly doped semiconductors  (often referred to as ‘designer' metals), can be used as plasmonic materials\cite{Law2012Mid-infraredMetals,Hoffman2007NegativeMetamaterials}. High quality, single-crystal infrared plasmonic doped semiconductors can be grown by molecular beam epitaxy (MBE), with the exquisite control of layer thickness and free carrier concentration afforded by epitaxial growth. 

\begin{figure*}
\includegraphics[width=\columnwidth]{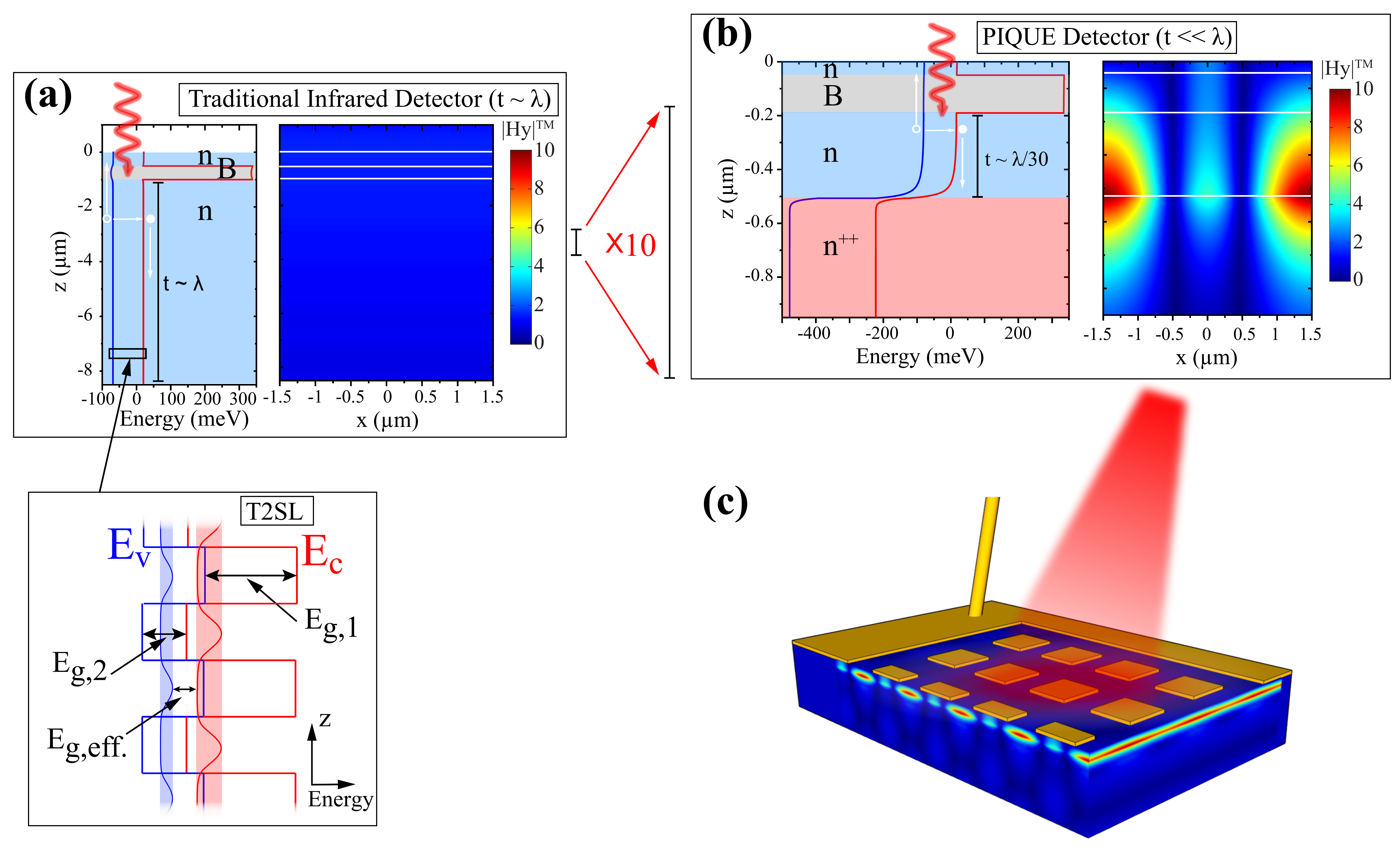}
  \caption{Depiction of the relative length scale of (a) a traditional LWIR detector architecture compared to (b) our Plasmonic Infrared Quantum-engineered Ultra-thin Epitaxial (PIQUE) Detector. Both (a) and (b) show the detector effective bandstructure as well as the field profile $|Hy|^{TM}$ {in the xz plane} for incident light {(with electric field polarized in the x-direction)} at the resonant frequency of the PIQUE detector ($\lambda=10.7~\mu m$). A significantly enhanced field amplitude is observed in the PIQUE detector, resulting in strong absorption in the ultra-subwavelength thickness ($\ll \lambda_0$) absorber region.  State-of-the-art performance is enabled by integrating the quantum engineered type-II superlattice absorbers (expanded view from (a)) in an nBn detector architecture with the strongly confined  SPP mode at the interface between the $n^{++}$ semiconductor and the detector stack. (c) Isometric schematic of the PIQUE detector with overlaid field profiles in the absorber region.}
  \label{fig:1}
\end{figure*}

Moreover, doped semiconductor plasmonic materials can be seamlessly integrated with MBE-grown quantum engineered optoelectronic active regions \cite{Nordin2020,Nordin2020All-EpitaxialResponsivity}. Notably, the class of semiconductor heterostructures known as type-II superlattices (T2SLs), consisting of alternating layers of III-V semiconductor alloys with type-II band offsets, allow for engineering of effective bandgap energies across the infrared, including energies well below the bandgap of the superlattice's constituent materials (Fig. \ref{fig:1}). T2SLs are of particular interest for infrared detectors, due to suppressed Auger recombination\cite{Grein1992MinoritySuperlattices}, and have been integrated into a variety of infrared detector architectures. {In particular, LWIR T2SLs have been used in so-called "bariode" devices such as the nBn detector, which consists of an n-doped absorber (n), a wide bandgap barrier (B) and a thin, n-doped contact layer (n).  Photocurrent is carried by photo-excited holes (minority carriers), while the wide bandgap barrier blocks transport of the majority carrier electrons, substantially reducing dark current and dramatically improving detector performance \cite{Maimon2006NBnTemperature, Ting2009ADetector}}.  The ability to directly integrate these quantum engineered infrared active devices with plasmonic materials, in an all-epitaxial system, provides a complete materials framework for the design of scalable sub-diffraction-limit detectors (or more generally, optoelectronic devices), in a wavelength range of vital importance for a host of sensing and imaging applications. 

{In order to quantify the benefits associated with sub-diffraction-limit thickness detector architectures, we consider the figure of merit specific detectivity ($D^*$), used to quantify infrared photodetector performance, across detector architectures}. Assuming uniform thermal generation and recombination rates, specific detectivity can be expressed as\cite{AntoniRogalski2010InfraredEdition},
\begin{equation}
D^* = \frac{\lambda_0\eta}{ hc \sqrt{2t(G + R)}}\frac{A_o}{A_e},
\label{eq:1}
\end{equation}
where $\lambda_0$ is the wavelength of light, $\eta$ is the external quantum efficiency of the detector (EQE, defined as the ratio of collected charge carriers to incident photons), $h$ is Planck's constant, $c$ is the speed of light, $t$ is the detector absorber thickness, $G$ and $R$ are the absorber generation and recombination rates respectively, and $A_o$ and $A_e$ are the device optical and electrical area, respectively. Optimizing detector performance, for a given wavelength and lateral geometry ($A_o$ and $A_e$), requires maximizing $\eta/\sqrt{t(G+R)}$; achieving high quantum efficiency while minimizing absorber thickness or the generation and/or recombination rates of carriers. HgCdTe (MCT) has been the material system of choice for infrared detectors for the last sixty years, due to its high quantum efficiency and small generation and recombination rates. Recent results \cite{Law19} have shown that careful heterostructure engineering can further reduce the generation and recombination rates of MCT detectors and therefore substantially lower their dark current. Although these results are encouraging, there is still interest in supplanting MCT due to environmental concerns associated with Hg and Cd\cite{EUDirective}. {III-V semiconductor-based detectors have been identified as a candidate replacement for MCT, as these detectors benefit from more uniform growth, widely used commercial substrates, and substantially easier fabrication\cite{Palaferri2018Room-temperatureReceivers}. However, some of the most promising III-V semiconductor-based detector candidates, quantum dot or quantum well infrared photodetectors (QDIPs and QWIPs, respectively), have struggled to realize EQE within an order of magnitude of MCT detector EQE\cite{Campbell2007Quantum-dotPhotodetectors,Barve2010SystematicPhotodetectors,Lu2007Temperature-dependentPhotodetector}. T2SL detectors on the other hand have struggled to achieve dark currents less than an order of magnitude above the longstanding MCT heuristic for dark current, Rule 07 \cite{Tennant2008MBEHeuristic}. In either case, the sub-optimal EQE or alternatively, dark current, ensures, for the III-V semiconductor-based detectors, $D^*$’s well below those of MCT.} In order to compete with existing MCT technology, next-generation IR detectors must be able to minimize detector thickness, or reduce generation/recombination rates, while maintaining the EQE typical of existing, wavelength-scale-thickness detectors and do so in an environmentally manageable material system. To this end, leveraging plasmonic modes for strong light confinement in detector absorbing regions allows for the decoupling of detector thickness and absorption efficiency, and provides a path towards high quantum efficiencies in ultra-thin absorbers, and therefore superior $D^*$. 

In this work we design, grow, fabricate, and characterize all-epitaxial, monolithically integrated and sub-diffraction-limit-thickness LWIR detectors leveraging surface plasmon-polaritons (SPPs) at a quantum engineered absorber/‘designer’ metal interface.  We dub these devices Plasmonic Infrared Quantum-engineered Ultra-thin Epitaxial (PIQUE) Detectors, and show that the PIQUE device architecture can dramatically reduce deleterious detector properties such as dark current, resulting in low-noise {operation}, while maintaining state-of-the-art detector response, all in a mature III-V semiconductor material system. The temperature-dependent electrical and optical properties of fabricated detectors are characterized experimentally, and the optical response compared to rigorous coupled wave analysis and finite element method simulations with excellent agreement. The PIQUE detectors presented outperform state-of-the-art commercial detectors, {both MCT and type-II superlattice detectors}\cite{VIGO, VIGOT2SL}, operate at temperatures compatible with thermo-electric coolers (TECs), and have the potential to extend efficient LWIR detector operation to room temperatures. {By monolithically integrating semiconductor plasmonic materials with sub-diffraction-limit thickness optoelectronic active regions, this work demonstrates, for the first time, a scalable all-epitaxial plasmonic optoelectronic architecture, and, importantly, a new approach to the design and development of mid-infrared optoelectronics}.

\section*{Results}
\subsection*{Optical Properties and External Quantum Efficiency}
The PIQUE detectors are grown by MBE on GaSb substrates, and consist of a type-II superlattice nBn detector with absorber/barrier/contact thicknesses of 311/146/46 nm above an epitaxially-grown $n^{++}$ semiconductor ‘designer’ metal. {Figure \ref{fig:1}(b) shows the growth stack and simulated bandstructure for the PIQUE detector. A 2D metal grating is patterned on the detector surface [as shown schematically in Fig. \ref{fig:1}(c)] to enable polarization-independent coupling into the SPP modes supported at the absorber/$n^{++}$ interface. Detector dimensions ($n^{++}$, contact, barrier, and absorber layer thicknesses, grating period and duty cycle) were optimized to give ultra-thin detectors (thus resulting in low dark currents), which could still achieve the integrated absorption/EQE (in the 8 - \SI{13}{\um} LWIR) of traditional, wavelength-scale thickness T2SL detectors\cite{Rogalski2019Type-IIPhotodiodes,Rhiger2011PerformanceHgCdTe, Rogalski2017InAs/GaSbProspect}. Epitaxial growth, detector fabrication, material characterization and optical properties are described in further detail in Supplement 1, Section 1.}

{The field plot ($|H_y|^{TM}$) in Fig. \ref{fig:1}(b) depicts the optical mode excited at resonant coupling into our detector structure. This mode is identified as a SPP mode, clearly bound to, and with peak intensity at, the absorber/$n^{++}$ interface.  Below the interface, the SPP decays rapidly in the plasmonic $n^{++}$ layer.  Above the interface, the SPP mode decays into the (dielectric) detector stack such that upper half of the mode is almost entirely confined in the ultra-thin (t = \SI{311}{\nm}) absorber layer. This tight mode confinement means very little field is present in the 2D metal grating and thus there is negligible loss in the gratings; the simulated loss/absorption in each layer is provided in Supplement 1, Section 1.  While the patterned 2D grating on the detector surface provides the in-plane momentum for coupling into the SPP modes, the strong mode confinement (and thus overlap of the optical field with the detector absorber) comes from the plasmonic mode supported by the $n^{++}$ ground-plane.  In Supplement 1, Section 1, we demonstrate that the 2D metal grating can be replaced with patterned highly doped (plasmonic) or undoped (dielectric) semiconductor material with minimal effect on the detector's optical properties (or the optical mode excited), underscoring that the plasmonic nature of the device comes from the doped semiconductor ground-plane, and not the metallic grating.}  

\begin{figure*}
\includegraphics[scale=0.3]{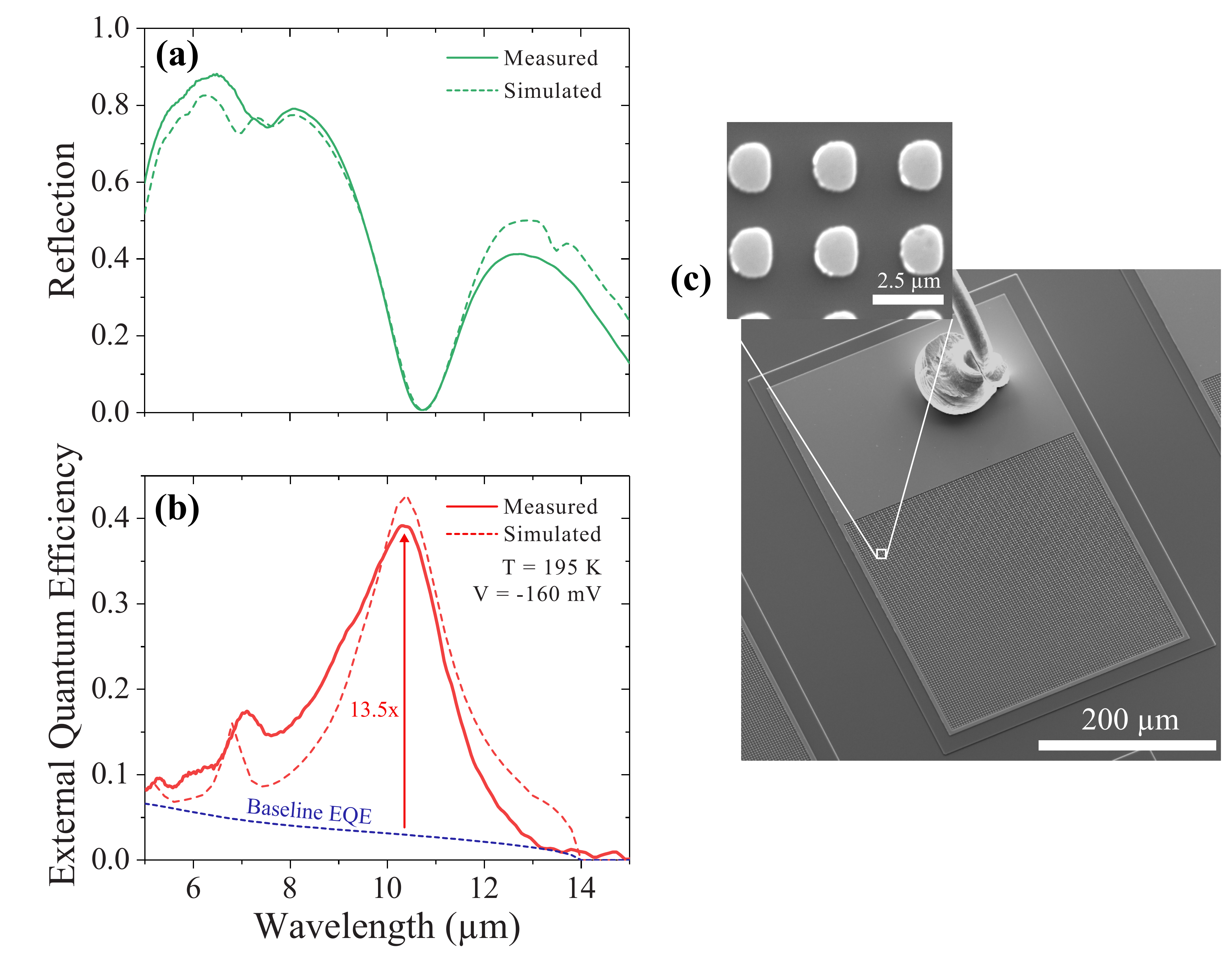}
  \caption{(a) Experimental (solid) and RCWA-simulated (dashed) room-temperature reflection spectra from the fabricated detector structure (b) External quantum efficiency of the PIQUE detector operating at \SI{195}{\kelvin} with a low reverse bias of \SI{-160}{\milli\volt}. The red solid line is the measured external quantum efficiency and the dashed line is the simulated external quantum efficiency using COMSOL multiphysics. The navy dashed line is the simulated external quantum efficiency of a detector with the same absorber thickness, but no plasmonic layer or Au {2D metal grating}. (c) Scanning electron micrograph of a fabricated PIQUE detector and an expanded view of the {2D metal grating}.}
  \label{fig:2}
\end{figure*}

The experimental room temperature reflection spectrum of a fabricated detector is shown in Fig. \ref{fig:2}(a), where the prominent dip in reflection at $\lambda_0=\SI{10.7}{\um}$ corresponds to coupling into SPP modes, as can be seen from the simulated field profile of the structure shown in Fig. \ref{fig:1}(b). {In the LWIR there is little transmission through the $n^{++}$ ground-plane and no diffraction from the $\Lambda=\SI{3}{\um}$ grating, thus reflection ($R$) in the LWIR can be related to total absorption ($A$) by the simple relation $A\simeq1-R$.  While a significant amount of this absorption occurs in the detector absorber (where it will be converted to photocurrent), some fraction will also be absorbed by the $n^{++}$ ground-plane and (to a much lesser extent) the gratings (light which will not be converted to photocurrent).  Supplement 1, Section 1 shows the relative absorption of the various components of our structure, as a function of wavelength.}

The resulting spectral EQE for a detector with {2D metal grating element width} \SI{1.5}{\um} and period \SI{3}{\um} is shown in Fig. \ref{fig:2}(b) and compared to finite element simulations of EQE for the same device with (red dashed) and without (blue dashed) $n^{++}$ ground-plane and {2D grating}.  Strong response is observed across the LWIR, with a peak EQE at $\lambda_0=\SI{10.4}{\um}$ of $\sim39\%$, a factor of $13.5\times$ enhancement in response when compared to the same detector structure's absorption, simulated without the underlying plasmonic material and {2D metal grating}. {In Supplement 1, Section 1, we also show the simulated absorption (and EQE) for the same nBn detector without the patterned 2D grating but with the underlying $n^{++}$ ground-plane, with the 2D grating and no $n^{++}$ ground-plane, and again, without either the 2D grating or the $n^{++}$ ground-plane.  There is negligible response in the detector structure with only the patterned metal, clearly demonstrating that patterned metal is not acting as an antenna array, but as a grating coupler.} The measured spectral response of the fabricated detector mirrors the absorption feature in Fig. \ref{fig:2}(a), with the slight spectral shift coming from the difference in sample temperature between the two experiments.  Notably, for the same T2SL material, we calculate that a detector with absorber thickness of $t\sim\SI{7.5}{\um}$ would be required to achieve the same EQE at $\lambda_0=\SI{10.4}{\um}$, optimistically assuming perfect photo-excited charge carrier collection efficiency. 

\subsection*{Temperature Dependent Responsivity and Dark Current-Voltage}
\begin{figure*}
\includegraphics[scale=0.3]{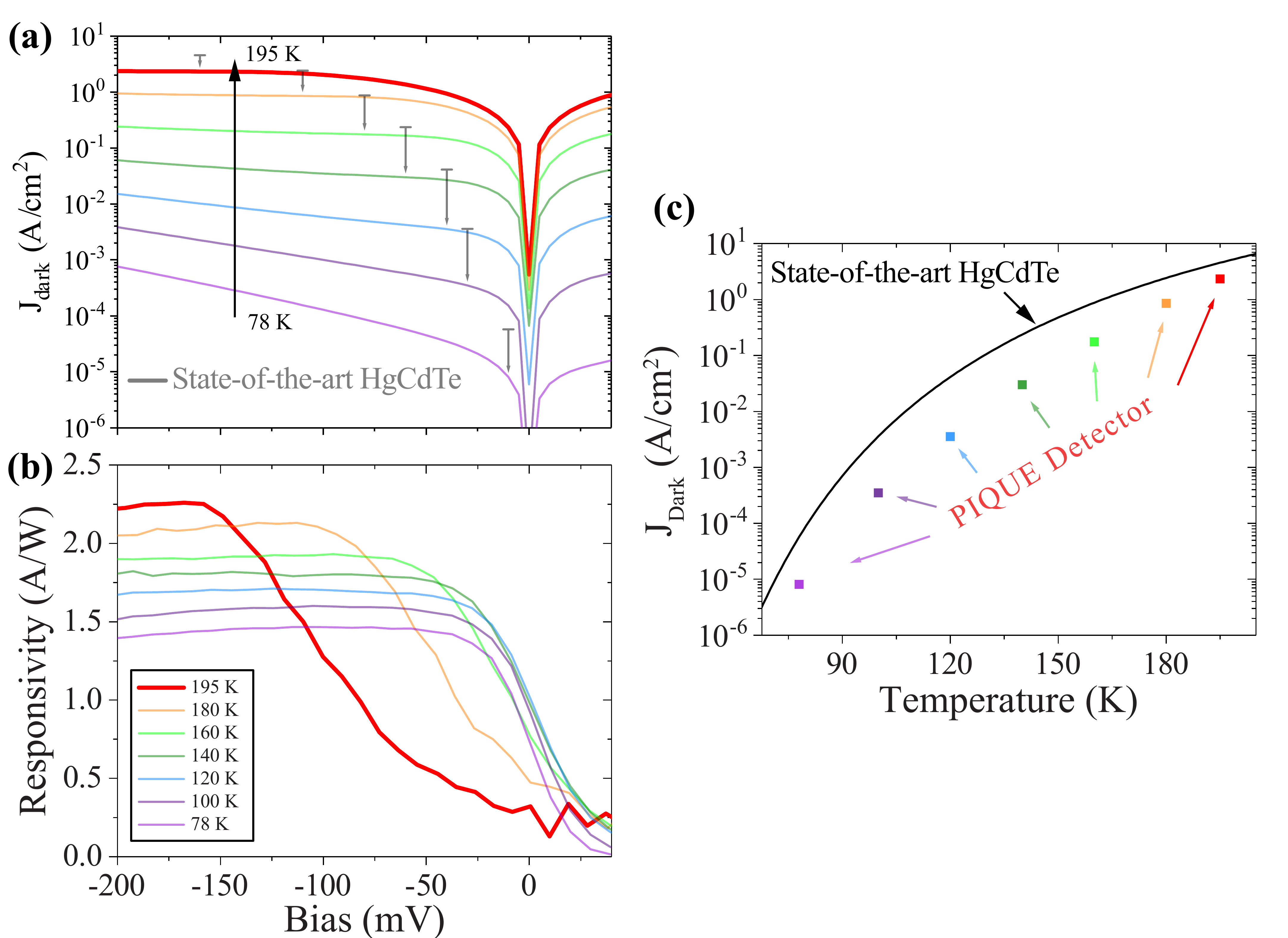}
  \caption{(a) Dark current, as a function of applied bias, of the ultra-thin detector for temperatures between $78-195~K$. The dash above each line is the commensurate Rule 07 dark current for each temperature, positioned at the bias of peak specific detectivity of the PIQUE detector.  (b) Bias dependent responsivity of the ultra-thin detectors taken at $\lambda_0=$\SI{9.46}{\um} as a function of temperature. As expected for a nBn detector, responsivity saturates  at low voltages (\SI{-80}{\milli\volt} at T = 78 K and \SI{-160}{\milli\volt} at T = 195 K) and remains constant with increasing bias. A slight increase in detector responsivity is observed with increasing temperature, resulting from the red-shifting of the T2SL band gap. (c) PIQUE detector dark current as a function of temperature plotted with the Rule 07 heuristic for a detector with $\lambda_{co} =$ \SI{11.42}{\um}. We choose the comparison detector cut-off wavelength to be at 50\% of the PIQUE's peak EQE (potentially significantly under-estimating the actual Rule 07 dark current). As can be seen, from 78 K to 195 K the plasmonic detector dark current out performs an ideal MCT detector (while maintaining comparable EQE).}
  \label{fig:3}
\end{figure*}

The ability to detect LWIR light, in a $\sim \lambda_0 / 33$ thickness detector, with effectively the same EQE as a wavelength scale thickness detector, has significant implications for our detector performance metrics given the relationship between $D^*$ and detector thickness from Eq. \ref{eq:1}. Figure \ref{fig:3}(a) shows the measured detector dark current (solid lines) for temperatures from 78 K to 195 K, compared to the dark current for a HgCdTe detector at the same temperatures (dashes), the latter calculated from the Rule 07 heuristic\cite{Tennant2008MBEHeuristic} which provides a rule-of-thumb estimate for state-of-the-art IR detectors. To accurately compare our fabricated detector dark current, at each temperature, to the Rule 07 dark current, the appropriate operational bias of the plasmonic detectors must be determined. Figure \ref{fig:3}(b) shows the temperature-dependent detector responsivity (at $\lambda_0=\SI{9.46}{\um}$) as a function of applied bias, with the strong saturation at low applied biases characteristic of nBn detectors and essential for compatibility with focal-plane-array (FPA) read-out circuitry.  We choose, for each temperature, an operational bias corresponding to peak specific detectivity.  The dark currents for these applied biases, as a function of temperature, are plotted in Fig. \ref{fig:3}(c), along with the Rule 07 dark current (which is typically already an order of magnitude less than the dark current of a traditional, wavelength-scale, T2SL nBn detectors \cite{Rogalski2019Type-IIPhotodiodes,Rhiger2011PerformanceHgCdTe, Rogalski2017InAs/GaSbProspect}).  Figure \ref{fig:3}(c) clearly demonstrates that the PIQUE detector dark current, for the full temperature range (78 K - 195 K) is substantially lower than the dark-current for state-of-the-art MCT LWIR detectors. Specifically, our dark current is lowest, relative to Rule 07, at 110 K with a dark current 12.8 $\times$ lower than Rule 07. However at elevated temperatures, specifically 195 K, we observe dark currents only 1.94 $\times$ lower than Rule 07. We attribute our detectors' dark current converging towards Rule 07 at elevated temperatures to the conservative choice of cutoff wavelength we use for the Rule 07 heuristic (choosing the wavelength where we observe 50\% of maximum EQE). In reality, our detector's T2SL absorber cutoff wavelength has a strong temperature dependence, whereas the optical enhancement associated with the SPP is largely temperature independent. Thus, we expect if we used the bandgap of the T2SL for our Rule 07 curve, rather than 50\% max EQE, we would maintain dark currents at least an order of magnitude lower than Rule 07 across the entire temperature range shown. 

\subsection*{Detector Performance}
\begin{figure*}
\begin{center}
    
\includegraphics[scale=0.3]{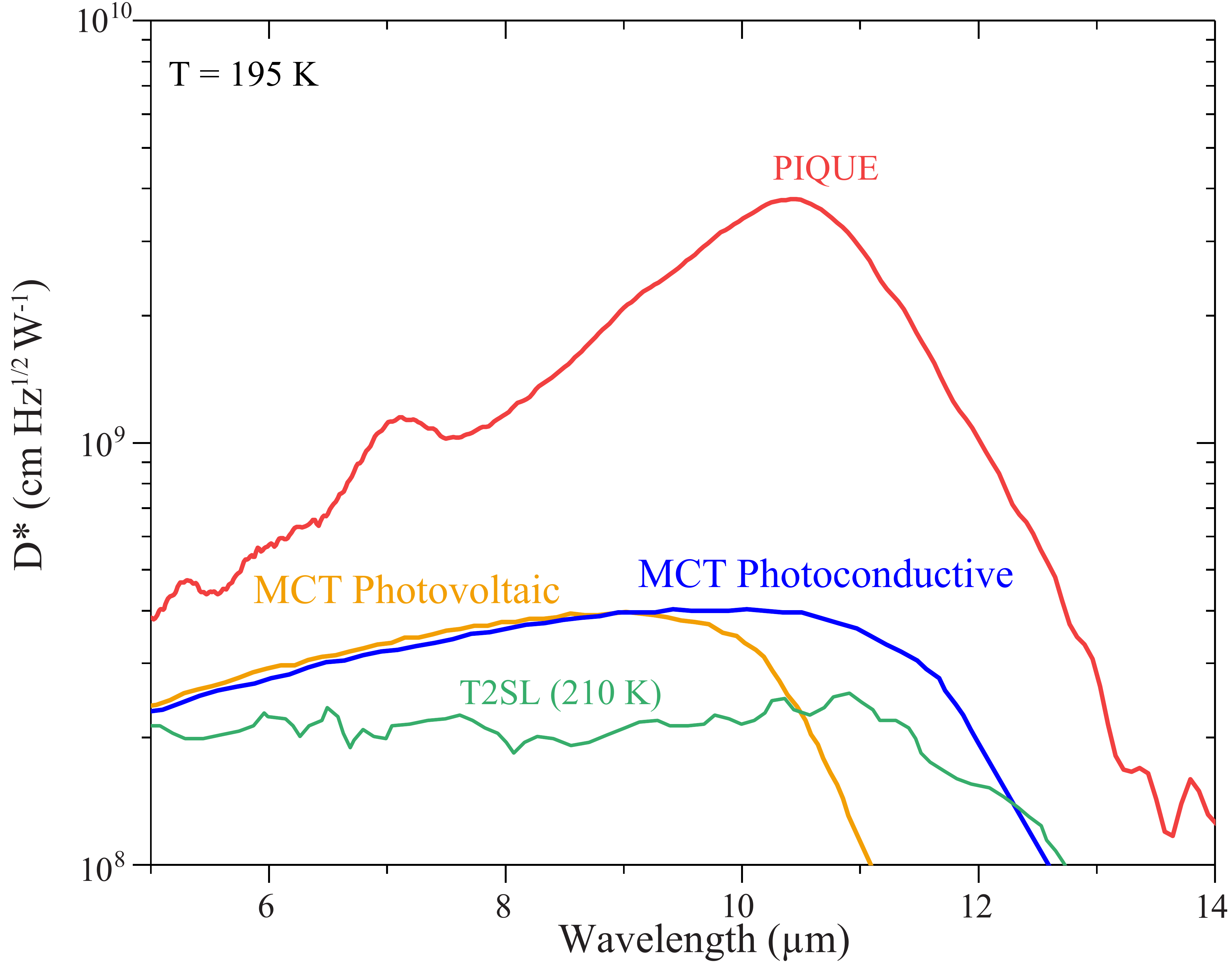}
  \caption{(a)  Spectral specific detectivity of the PIQUE detector and commercial LWIR detectors at comparable temperatures. As can be clearly seen, PIQUE detectors at 195 K substantially outperform commercial HgCdTe at 195 K and T2SL detectors at 210 K. }
  \label{fig:4}
  \end{center}

\end{figure*}

The significantly reduced dark currents achievable in  PIQUE detectors has a dramatic impact on the detectors' $D^*$.  Fig. \ref{fig:4} plots detector performance ($D^*$) as a function of wavelength, for our PIQUE detectors, as well as state-of-the-art commercially-available LWIR detectors.  As can be observed from this plot, the PIQUE detector dramatically outperforms existing LWIR detectors, due to the significant improvement in dark current associated with the $\sim\lambda_0/33$ absorber thickness. This improvement in dark current can be realized without any degradation in EQE, a result of the strong field enhancement achieved in the plasmonic structure. The PIQUE detector exhibits peak $D^*\approx\SI{4E9}{\cm \Hz^{1/2} \W^{-1}}$ at $T=195~$K, where the PIQUE detector outperforms both commercial T2SL\cite{VIGOT2SL} and MCT\cite{VIGO}  detectors. However, {The PIQUE detectors presented here are single-element, front-side illuminated (as are the comparison commercial detectors), and not not well-suited for the substrate side illumination required for most FPA designs. Future detector designs will look to reverse the order of growth (nBn detector, then $n^{++}$ plasmonic layer, to enable coupling to SPPs from substrate-side illumination and thus FPA configurations.}

The single greatest challenge for LWIR detectors is the demonstration of high performance at elevated temperatures. The detector performance of any LWIR photodetector degrades dramatically with increasing temperature, with $D^*$ typically decreasing by two or more orders of magnitude as operating temperature increases from liquid nitrogen (77K) to room ($\sim300$K) temperature.  This degradation is almost completely a result of the exponential increase in dark current density, entirely expected for a semiconductor device whose effective band gap energy is essentially the energy of the peak blackbody emission at detector operating temperatures near 300 K.  Decreasing the dark current of a LWIR detector, for a given detector design, can only be achieved by drastically reducing the detector volume, which for traditional detector architectures incurs a severe responsivity penalty. Not only do our PIQUE detectors provide a $\sim20\times$ reduction in absorber thickness (compared to state-of-the-art LWIR detectors), but they achieve this reduction in absorber thickness without any appreciable degradation of detector responsivity. These results and the presented PIQUE detector architecture offer a realistic path towards high performance room temperature LWIR detectors, in many ways the grail of LWIR detector work. {The all-epitaxial nature of our device architecture is the key element enabling the demonstration of high-performance plasmonic optoelectronic devices. At shorter wavelengths (near-IR/vis), the thickness of a comparable interband diode detector, scaled for wavelength, would be $\sim\SI{50}{\nm}$. This $\sim\SI{50}{\nm}$ would need to include both the $n$-doped, $p$-doped, and depletion regions of the diode, not to mention $n$- and $p$-type contact layers. In addition, the structure would have to be grown over a plasmonic film without defects and dislocations, as defects and dislocations would degrade the device’s optical and electrical performance. Even were all this possible, the high absorption coefficients and low $G$ and $R$ coefficients of the wide-bandgap near-IR/vis semiconductors leave much less to be gained by leveraging thin absorber regions. Thus, the monolithic, all-epitaxial integration of the PIQUE detectors offers a unique and novel opportunity to explore plasmonic optoelectronic device architectures while simultaneously realizing significant performance enhancement over existing mid-infrared technologies.}     

\section*{Conclusion}
In summary, we demonstrate a long-wave infrared photodetector, leveraging a quantum engineered type-II superlattice absorber integrated into an nBn photodetector architecture, and embedded in a resonant epitaxial plasmonic structure.  The strong confinement of incident light achieved by coupling into SPP modes at the detector/highly doped semiconductor interface allows for an ultra-thin detector architecture without any loss of detector responsivity. {Moreover, the presented architecture is extremely versatile and could be implemented in other material systems and/or extended to longer wavelengths.}  We show detector dark currents 1.9 $\times$ below the Rule 07 heuristic, and peak external quantum efficiencies of $39\%$ (which would require thicknesses on the order of $>\SI{7.5}{\um}$ in a traditional detector architecture).  The presented detectors show specific detectivity above that of state-of-the-art commercial LWIR photodetectors, both MCT and T2SL. The PIQUE detectors presented in this work utilize an all-epitaxial plasmonic-optoelectronic device design with state-of-the-art performance demonstrating a scalable plasmonic architecture for real-world applications, capable of outperforming commercial detectors and offering a viable alternative to current, environmentally problematic and widely used, detector materials.

\begin{backmatter}
\bmsection{Acknowledgements}

DW and LN gratefully acknowledge support from the National Science Foundation (ECCS-1926187) and Lockheed Martin Co.. PP gratefully acknowledges support from the National Science Foundation CDCM MRSEC (DMR-1720595). AJM gratefully acknowledges support from the Defense Advanced Research Projects Agency (NASCENT, NLM program). This work was partly done at the Texas Nanofabrication Facility supported by NSF grant NNCI-1542159. 

\bmsection{Disclosures}
The authors declare that they have no competing interests.

\bmsection{Data Availability}
Data underlying the results presented in this paper are not publicly available at this time, but may be obtained from the authors upon reasonable request.

\bmsection{Code Availability}
The room-temperature reflection RCWA modeling was performed using a Matlab software package from V. A. Podolskiy’s research group, see \url{http://faculty.uml.edu/vpodolskiy/codes/index.html}  for an implementation. The numerical absorption modeling was performed using COMSOL multiphysics software.  The specific simulation files and or data related to either of these software packages or the code that analyses the experimentally measured data is available from the corresponding author upon reasonable request.

\bmsection{Supplemental Document}
See Supplement 1 for supporting content. In particular, epitaxial growth details, fabrication and characterization procedures, modeling and simulation descriptions, and x-ray diffraction spectra of the PIQUE detector material.  {Supplement 1 also provides a detailed description of the simulated spectrum in Fig. \ref{fig:2}(b), as well as additional simulations showing the fractional absorption in the different components of the detector, and the performance of detectors with/without gratings and $n^{++}$ ground-planes}.
\end{backmatter}

%%%%%%%%%%%%%%%%%%%%%%% References %%%%%%%%%%%%%%%%%%%%%%%%%

%%%%%%%%%% If using BibTeX:
\bibliography{references}

%%%%%%%%%% If preparing manually:
% \newcommand{\enquote}[1]{``#1''}

% \bibitem{Zhang:14}
% Y.~Zhang, S.~Qiao, L.~Sun, Q.~W. Shi, W.~Huang, L.~Li, and Z.~Yang,
%   \enquote{Photoinduced active terahertz metamaterials with nanostructured
%   vanadium dioxide film deposited by sol-gel method,}
%   {\protect\JournalTitle{Optics Express}} \textbf{22}, 11070--11078 (2014).

% {Optical Society}, \enquote{{OSA Publishing},}
%   \url{http://www.osapublishing.org}.

% \bibitem{FORSTER2007}
% P.~Forster, V.~Ramaswamy, P.~Artaxo, T.~Bernsten, R.~Betts, D.~Fahey,
%   J.~Haywood, J.~Lean, D.~Lowe, G.~Myhre, J.~Nganga, R.~Prinn, G.~Raga,
%   M.~Schulz, and R.~V. Dorland, \enquote{Changes in atmospheric consituents and
%   in radiative forcing,} in \enquote{Climate Change 2007: The Physical Science
%   Basis. Contribution of Working Group 1 to the Fourth assesment report of
%   Intergovernmental Panel on Climate Change,}  S.~Solomon, D.~Qin, M.~Manning,
%   Z.~Chen, M.~Marquis, K.~B. Averyt, M.~Tignor, and H.~L. Miler, eds.
%   (Cambridge University Press, 2007).

% \end{thebibliography}

\end{document}

\maketitle

\section{Optical Design}
The optical properties of the detector are modeled using rigorous coupled wave analysis (RCWA)\cite{FromAnalysis.} and finite element methods (COMSOL)\cite{COMSOL}. The $n^{++}$ plasmonic layer is well-described using the Drude formalism\cite{Law2012Mid-infraredMetals}, 

\begin{equation}
\epsilon_m(\omega)=\epsilon_b\left(1-\frac{\omega_p^2}{\omega^2+i\omega\gamma}\right)
,\label{eq:2}
\end{equation}

with plasma wavelength $\lambda=2\pi c/\omega_{p} =\SI{5.5}{\um}$ and scattering rate $\gamma =\SI{1e13}{}$ rad/s, extracted from reflection spectra taken on $n^{++}$ calibration samples consisting of only the heavily doped type-II superlattice (T2SL). The background permittivity, $\epsilon_b=12.56$ , is determined by a weighted average of the constituent alloys. The barrier superlattice is modeled as a lossless dielectric with permittivity $\epsilon=12.15$, also determined by the weighted average of the constituent alloys. The GaSb substrate is treated as lossless dielectric layer with constant permittivity $\epsilon=14.4$. 

\subsection*{Optical architecture}
\begin{figure*}[h!]
\begin{center}
\includegraphics[scale=0.19]{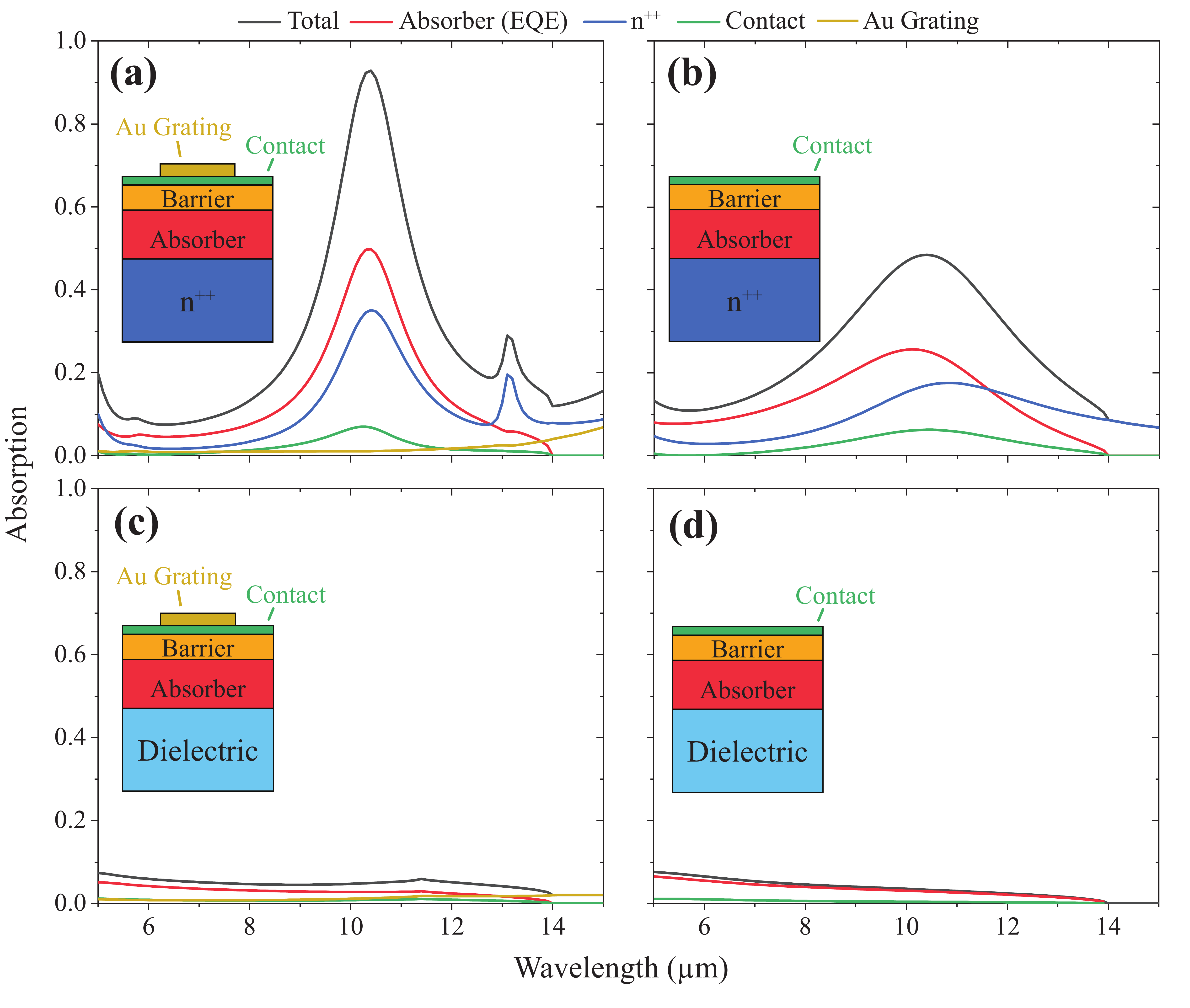}
  \caption{Simulated total absorption the absorption in each layer for four candidate detector architectures. (a) A nBn detector grown above a n$^{++}$ ground-plane and patterned with a grating coupler. (b) A nBn detector grown above a n$^{++}$ ground-plane with no surface patterning. (c) A nBn detector grown above a high-index dielectric ground-plane patterned with a grating coupler. (d) A nBn detector grown above a high-index dielectric ground-plane with no surface patterning. While both (a) and (b) show resonant behavior, only (a) is coupling to a surface plasmon-polariton, as this SPP mode requires the momentum imparted by the grating.}
  \label{fig:S1}
  \end{center}
\end{figure*}
To investigate the contributions of the n$^{++}$ ground-plane and metal grating coupler to our device performance, we simulated detector designs with both the n$^{++}$ layer and grating coupler [Fig. \ref{fig:S1}(a), our experimentally investigated design], with only the n$^{++}$ layer and no grating [Fig. \ref{fig:S1}(b)], with only the grating coupler and the n$^{++}$ layer replaced with a lossless high index ($\epsilon_d = 12.25$) dielectric substrate [Fig. \ref{fig:S1}(c)], and with no grating coupler  and the n$^{++}$ ground-plane replaced by the high-index substrate [Fig. \ref{fig:S1}(d)].  For each configuration, we plot the total simulated absorption (black), as well as the absorption in each of the layers: the detector absorber layer (red), the contact layer (green), the Au grating (gold), and the $n^{++}$ ground-plane (blue). The nBn detector and grating dimensions simulated are identical to the dimensions described in the main text.

Shown in Fig. \ref{fig:S1}(a) is the absorption for the detector design with both the n$^{++}$ layer and grating coupler. Strong absorption is clearly observed, peaking at \SI{10.4}{\um}, and the majority of that absorption is shared between the absorber layer and the plasmonic $n^{++}$ ground-plane, as would be expected for a surface plasmon-polariton (SPP) mode bound to this interface. While the light absorbed in the detector's absorber region can be collected as photocurrent, the (substantial fraction of) light absorbed in the n$^{++}$ will not contribute to the device photocurrent.  While such loss is inevitable in strongly confined plasmonic modes, the total absorption in the detector remains comparable to the absorption of traditional wavelength scale thickness long-wave infrared (LWIR) T2SLs\cite{Rogalski2019Type-IIPhotodiodes,Rhiger2011PerformanceHgCdTe, Rogalski2017InAs/GaSbProspect}. Notably, the metallic grating coupler has negligibly low absorption ($<1\%$), further confirming that the mode is primarily bound to the epitaxial $n^{++}$/detector interface, not the metallic grating coupler. 

Fig. \ref{fig:S1}(b) presents the absorption spectra for the device configuration where the grating coupler is removed but the n$^{++}$ plasmonic layer remains below the nBn detector. Here we observe that the total absorption is lower by a factor of two [when compared to the configuration of Fig. \ref{fig:S1}(a)], but still appreciable for such a thin active region. This absorption is associated with a leaky mode of the air-detector-n$^{++}$ stack. Such modes have been demonstrated\cite{Nordin2020,Nordin2020All-EpitaxialResponsivity}, and leveraged for enhanced detection, using all-epitaxial plasmonic materials.  However, though such structure include 'plasmonic' materials, they cannot be considered plasmonic devices as they do not leverage plasmonic modes (and the associated strong confinement achievable with such modes). Additionally, the leaky mode absorption observed in Fig. \ref{fig:S1}(b) requires an air or other low-index superstrate above the detector, making this architecture unrealistic for future substrate-side illumination detector designs.

Fig. \ref{fig:S1}(c) shows the absorption for the detector configuration with a grating coupler but no underlying n$^{++}$ ground-plane.  The simulation assumes the nBn detector is grown on a lossless, high-index semiconductor substrate with $\epsilon_d = 12.25$.  The total absorption for this configuration is extremely low, underscoring the critical role the epitaxial n$^{++}$ layer plays in our detector architecture. Lastly, shown in Fig. \ref{fig:S1}(d) is the detector configuration with neither the n$^{++}$ layer or the grating coupler; this is the simulated 'baseline EQE' spectrum plotted in Fig. 2(b) of the main text. As anticipated, the low absorption coefficient of the T2SL detector, and the lack of any surrounding photonic structure designed to concentrate the optical fields, results in very low EQE. Such a design would produce a low performance detector. 

\subsection*{Grating coupler}
\begin{figure*}[h!]
\begin{center}
\includegraphics[scale=0.1]{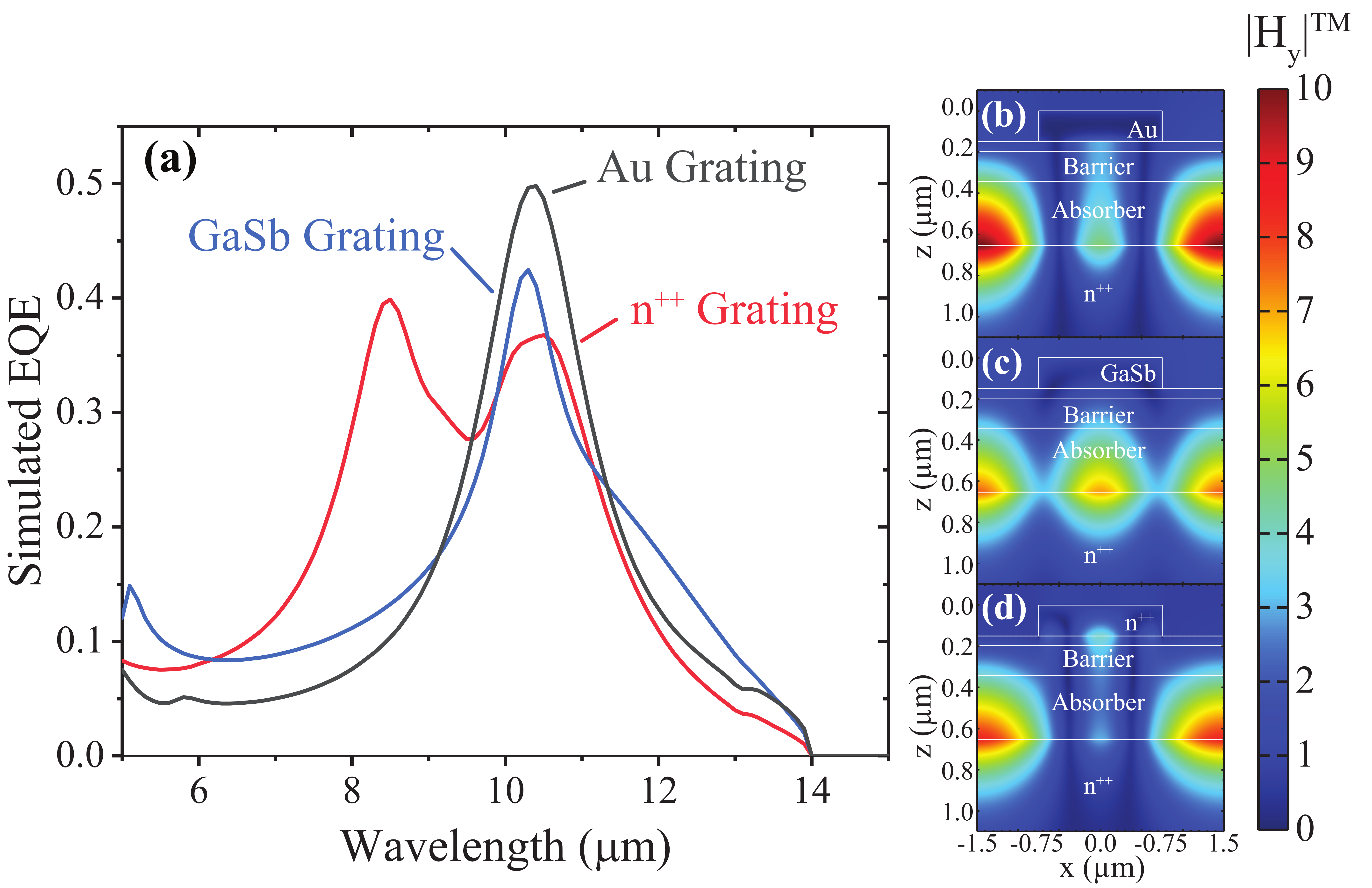}
  \caption{(a) Simulated EQE for plasmonic detectors with a metallic (Au) grating coupler (black), lossless dielectric (GaSb) grating coupler (blue), and a plasmonic (n$^{++}$) grating coupler (red). $|H_y|^{TM}$ field profile on resonance for the detector designs with (b) noble metal (Au) (c) GaSb and (d) n$^{++}$ grating couplers. All field profiles show a tightly bound interface mode, characteristic of a surface plasmon-polariton, at the absorber/n$^{++}$ ground-plane interface.}
  \label{fig:S2}
  \end{center}
\end{figure*}
In the previous section, which described the optical architecture of our detectors, we assumed a metallic grating coupler. However, grating couplers made of lossless dielectric or plasmonic materials can also provide the necessary momentum to allow coupling to the SPP mode at the nBn detector/n$^{++}$ ground-plane interface. We simulated the EQE for detector designs with three different grating materials: i) metal (Au), ii) plasmonic (identical to the n$^{++}$ ground-plane material), and iii) lossless high-index dielectric (GaSb, $\epsilon_d = 14.4$).  Importantly, the plasmonic n$^{++}$ or GaSb grating layers could be grown epitaxially, with the caveat that such a design would require an additional lithography and etch step to delineate the grating. Shown in Fig. \ref{fig:S2}(a) is the simulated EQE for a detector with each candidate grating coupler. The peak simulated EQE is largest for the Au grating coupler, followed by the GaSb grating, and then the n$^{++}$ grating coupler, which offers the lowest peak EQE.  Interestingly, the integrated response for the n$^{++}$ grating coupler is the largest of the three, a result of the additional shorter wavelength EQE peak observed at $\sim8.5~\mu$m. We associate this additional feature with coupling to SPP modes via the localized surface plasmon resonance supported by the individual n$^{++}$ grating layer elements. 

In addition to the simulated EQE for each grating material, we also show the simulated $|H_y|^{TM}$ field profiles on resonance for the Au [Fig. \ref{fig:S2}(b)], GaSb [Fig. \ref{fig:S2}(c)], and the n$^{++}$ grating [Fig. \ref{fig:S2}(d)]. As expected, the spatial profile of the excited modes are largely identical for the devices with the different grating couplers; the majority of the field is bound to the nBn detector/n$^{++}$ ground-plane interface. These simulations indicate that any grating coupler material (plasmonic, dielectric, or metallic) will facilitate coupling into the SPP mode at the  n$^{++}$ ground-plane/nBn detector interface. However, realizing the detector designs with the high-index dielectric or plasmonic grating couplers would require additional fabrication steps and yet provide lower performance (at least in peak EQE). These simulations reinforce the grating's role in coupling to the SPP mode, and indicate that metallic gratings are the optimal grating choice to minimize fabrication steps and maximize peak EQE.

\section{Epitaxial Growth}

\begin{figure*}
\begin{center}

\includegraphics[scale=0.4]{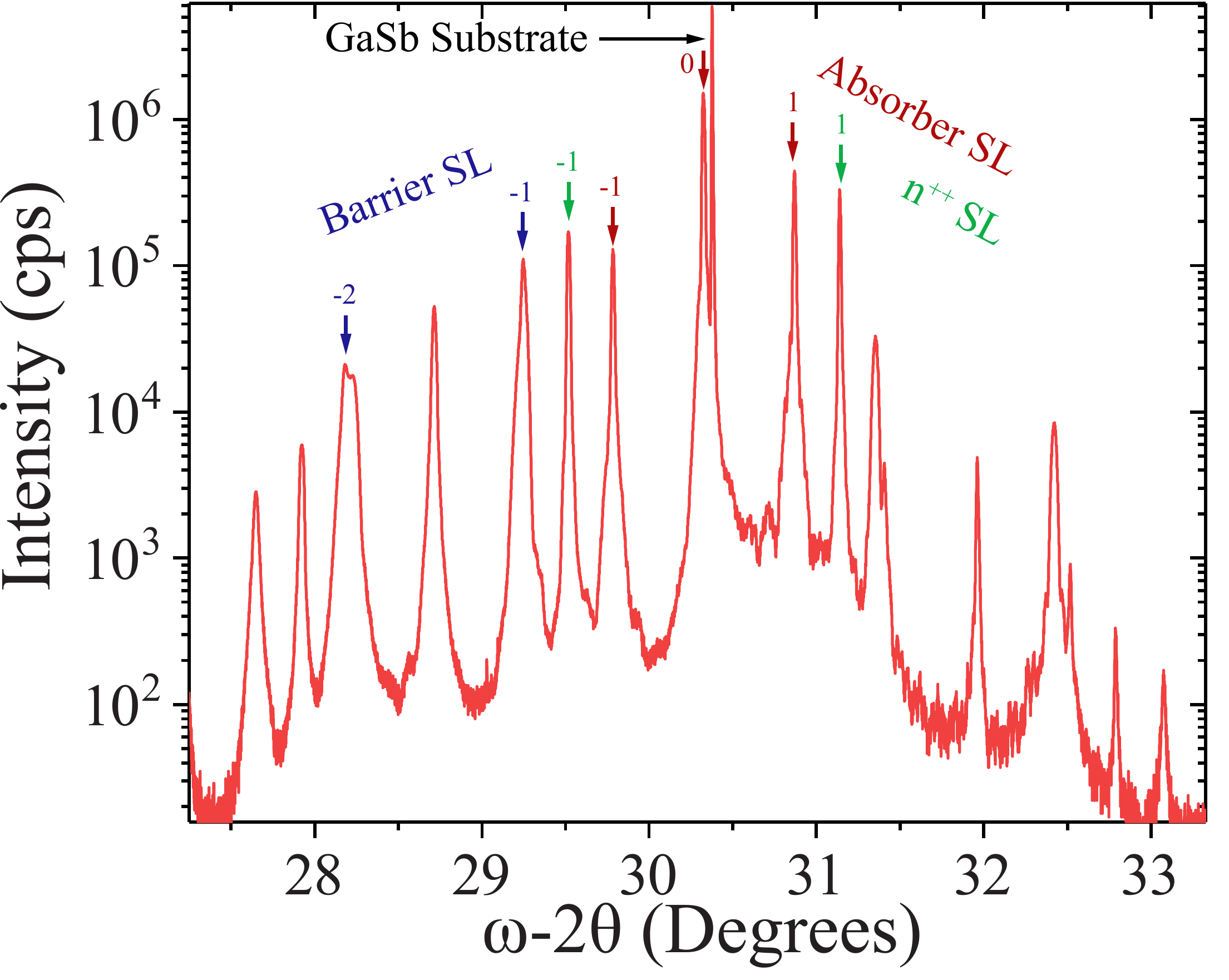}
  \caption{XRD (004) $\omega-2\theta$ scans of the PIQUE detector. The GaSb substrate peak, and several of the barrier SL and Absorber SL satellite peaks are identified. The plasmonic SL’s period is twice the absorber SL, so the satellite peaks overlap every other order. Satellite peak spacing allows us to identify the thickness of each constituent period giving thicknesses of 9.1 nm for the absorber SL, 4.88 nm for the barrier SL, and 6.2 nm for the plasmonic SL. The 0th order SL peak for the barrier/absorber/plasmonic SL all sit on top of each other at a slightly shallower angle than the GaSb substrate, meaning the SLs are slightly compressively strained (~2000 ppm) to the host substrate. }
  \label{fig:S3}
  \end{center}

\end{figure*}

Our structure is grown by molecular beam epitaxy in a Varian Gen-II system with effusion sources for gallium, indium, aluminum, and silicon, and valved cracker sources for arsenic and antimony. The PIQUE detector structure is grown on an n-type doped GaSb substrate, with the layer stack band structure shown in Fig. 1(b) of the main text. First, 750 nm  of a heavily doped ($n^{++}=\SI{5e19}{\cm^{-3}}$, with plasma wavelength $\lambda=2\pi c/\omega_{p} =\SI{5.5}{\um}$ and scattering rate $\gamma=\SI{1e13}{} $ rad/s) mid-wave infrared (MWIR) T2SL with a period of 20 ML, an alloy composition of  InAs/InAs\textsubscript{0.49}Sb\textsubscript{0.51}, and an InAs/InAsSb ratio of 16.5/3.5ML is grown, serving as the plasmonic ground-plane for the detector.   Note that while we use a T2SL as our plasmonic layer to simplify the lattice-matching to the layers above the plasmonic film, this T2SL can simply be treated as a Drude metal, as the combination of the MWIR cut-off and state filling from the high doping concentration (Burstein Moss shift) ensures that the only loss in the layer comes from the Drude scattering term ($\gamma$). Above the plasmonic ground-plane, we grow the nBn detector, which consists of a 311 nm  LWIR T2SL absorber with a period of 30 ML, an alloy composition of  InAs/InAs\textsubscript{0.49}Sb\textsubscript{0.51}, and an InAs/InAsSb ratio of 33/7ML, followed by a 146 nm barrier superlattice with period of 16 ML, and individual thickness and compositions of \textbf{1}/3/\textbf{1}/3/\textbf{1}/\underline{7} ML (\textbf{InAs}/AlAs\textsubscript{0.49}Sb\textsubscript{0.51}/\underline{InAs\textsubscript{0.49}Sb\textsubscript{0.51}}), and a 46 nm contact layer using the same superlattice as the absorber layer. The absorber and contact layer T2SLs are designed for a low temperature cut-off wavelength of $\lambda_{co}=11 \mu m$  The epitaxial structure geometry is measured using x-ray diffraction (XRD) in order to verify layer thickness and periods, see Fig. \ref{fig:S3} for XRD of the as-grown sample.  

\section{Detector Fabrication}

Following epitaxial growth, the as-grown material is patterned into mesas of dimensions \SI{340}{\um} x \SI{540}{\um} using UV photolithography and a wet etch.  The mesas are etched to a depth of 750 nm in order to provide electrical isolation between mesa top contacts.  The detector top and bottom contacts, as well as the 2D metal gratings are then defined using UV photolithography and a metallization and lift-off process.  The metal layer stack is 3 nm of Pd, 50 nm of Ti, 50 nm of Pt, and 100 nm of Au, which is a standard shallow contact for narrow bandgap T2SL materials. Detectors are left unpassivated, however following lift off the detectors are thoroughly cleaned in a bath of 90 °C AZ KWIKSTRIP followed by sonication in organic solvents (acetone, methanol, and isopropyl alcohol). 

\section{Detector Characterization}
The optical properties of the fabricated detector (Fig. 2 of the main text) are measured by infrared reflection spectroscopy, using a Bruker v80V Fourier transform infrared (FTIR) spectrometer coupled to a mid-IR microscope. The detectors are then mounted onto leadless chip carriers in a custom designed temperature-controlled cryostat with ZnSe windows.  Detector dark current as a function of applied bias (J-V) is measured using a Keithley 2460 low-noise source-meter. Dark measurements were taken with the detector blocked by a copper shield thermally connected to the cryostat cold-finger.
Fabricated detector response is measured using the FTIR spectrometer.  Light from the FTIR's internal source is focused, using a 2" diameter 6" focal length parabolic mirror, onto the detector, which is held in the same cryostat used for the J-V measurements. The measured spectral response of the detectors is normalized to the response of a spectrally flat pyroelectric detector in the same optical set-up, in order to remove the spectral weighting associated with the FTIR beamsplitter and the FTIR's internal blackbody source.  In order to convert the normalized spectral response to a responsivity, the detector response to a calibrated 673.15 K blackbody source is measured at $\lambda_{bp}=\SI{9.46}{\um}$ using an infrared bandpass filter.  Measuring the detector response at a fixed wavelength provides a scaling factor to convert the spectral response to a spectral responsivity in A/W.  Specific detectivity is then calculated using the detector dark current and responsivity in the following expression,

\begin{equation}
    D^* = R_i  \sqrt{
    \frac{A}{2 q \abs{I} + 4 k_b T/R}
    }
    \hspace{1mm},
    \label{eq:1}
\end{equation}
where $R_i$  is the measured responsivity, $I$ is the detector dark current,  $A$ is the mesa area, $q$ is the electronic charge, $k_b$ is Boltzmann's constant, $T$ is the detector temperature, and $R$ is the dynamic resistance. 

\section{Experimental Models}
\subsection*{Reflection model}
For the room temperature reflection RCWA model the simulated spectrum is an average of both the TM and TE polarization and the angular spread of the 36 x Cassegrain objective, 15 to 30 degrees, used to measure the sample. The absorber and contact layers are modeled with a simple $\alpha(E)=\alpha_0 \sqrt{E-E_g }$ dependence, where $\alpha_0=\SI{8000}{\cm^{-1}}$ and  $E_g=52.81$ meV $(\lambda_0=\SI{23.46}{\um})$. The room temperature bandgap is calculated using an 8-band k.p model. 
\begin{figure*}
\begin{center}
\includegraphics[scale=0.35]{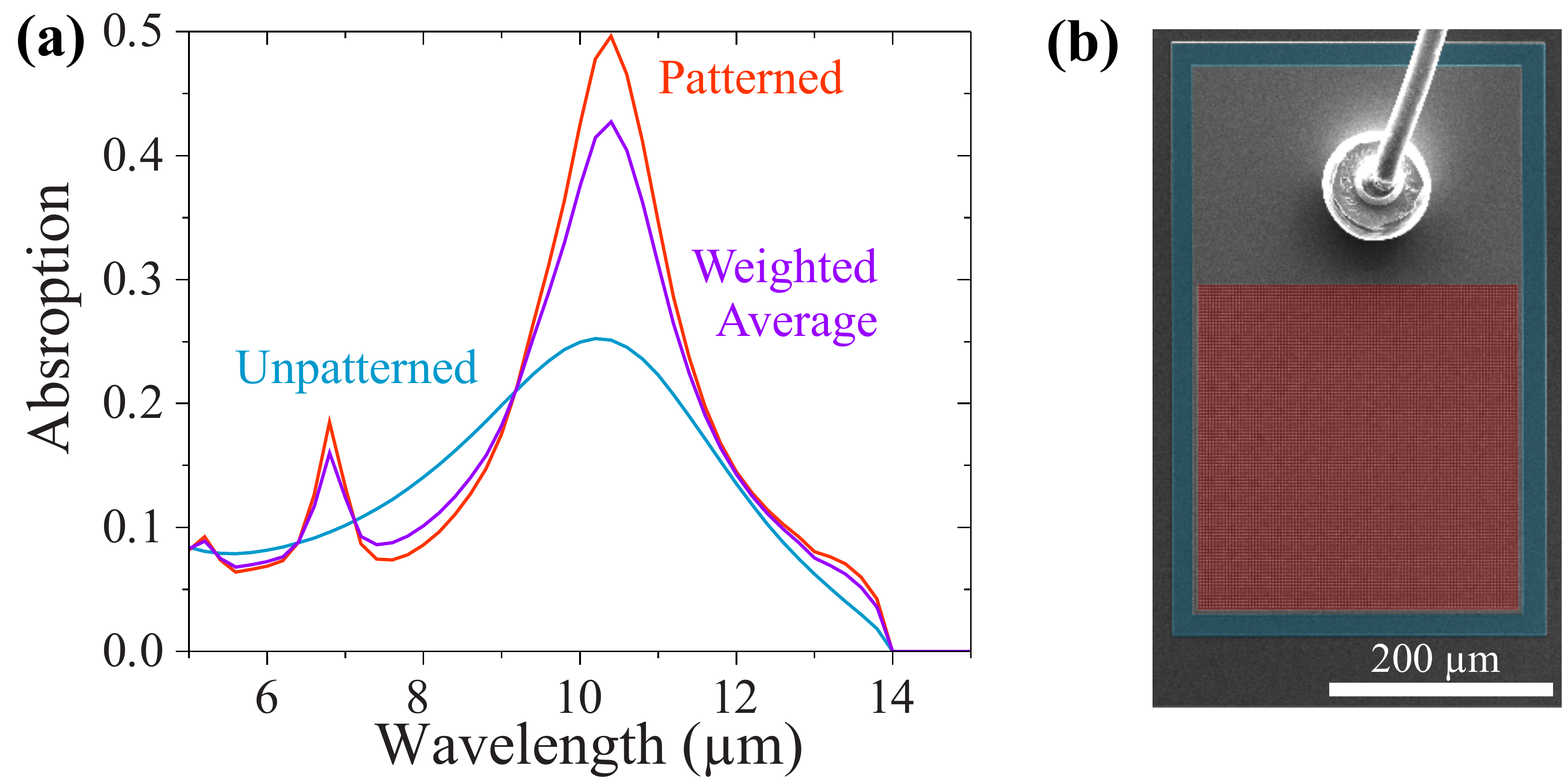}
  \caption{(a) COMSOL simulations for the absorption in the detector in patterned and unpatterned regions of the mesas. The simulated detector response presented in Figure 2b of the manuscript text is the weighted average, weighted by area, shown in purple.  (b) Scanning electron microscope image of a device mesa with false coloring to show the unpatterned (blue) and patterned (red) regions of the device mesas. }
  \label{fig:S4}
  \end{center}
\end{figure*}
\subsection*{EQE model}
For the 195 K absorption model COMSOL's wave optics module is used. The absorber and contact layers are modeled using the same simple $\alpha(E)=\alpha_0 \sqrt{E-E_g }$ dependence, where $\alpha_0=\SI{8000}{\cm^{-1}}$  and $E_g=88.5$ meV $(\lambda_0=\SI{14}{\um})$. To fully capture the measured spectrum the weighted average of the unpatterned region near the edge of the mesas [with absorption given by Fig. \ref{fig:S1}(b)] and the patterned region in the center of the mesas [with absorption given by Fig. \ref{fig:S1}(a)] is computed. Shown in Fig. \ref{fig:S4} is the schematic of the weighted averaging procedure. 

\bibliography{references}